\documentclass[runningheads]{llncs}
\usepackage{graphicx}
\usepackage{subfigure}
\usepackage{amsmath}
\usepackage{cases}
\usepackage{booktabs}
\usepackage{multirow}
\usepackage{amsfonts}

\makeatletter
\newcommand{\printfnsymbol}[1]{%
	\textsuperscript{\@fnsymbol{#1}}%
}
\makeatother
\begin{document}
\title{Label Refinement with an Iterative Generative Adversarial Network for Boosting Retinal Vessel Segmentation}
%
%
\author{
	Yunqiao Yang\inst{1}\thanks{equal contribution}
	\and
	Zhiwei Wang\inst{2}\printfnsymbol{1}
	\and
	Jingen Liu\inst{3}
	\and
	Kwang-Ting Cheng\inst{2}
	\and
	Xin Yang\inst{1}
}
%
%
\institute{Huazhong University of Science and Technology, China
	\and
	Hong Kong University of Science and Technology, Hong Kong
	\and
	JD AI Research, Mountain View, USA 
}

%
%
\maketitle      
\begin{abstract}
State-of-the-art methods for retinal vessel segmentation mainly rely on manually labeled vessels as the ground truth for supervised training. The quality of manual labels plays an essential role in the segmentation accuracy, while in practice it could vary a lot and in turn could substantially mislead the training process and limit the segmentation accuracy. This paper aims to “purify” any comprehensive training set, which consists of data annotated by various observers, via refining low-quality manual labels in the dataset. To this end, we have developed a novel label refinement method based on an iterative generative adversarial network (GAN). Our iterative GAN is trained based on a set of high-quality patches (i.e. with consistent manual labels among different observers) and low-quality patches with noisy manual vessel labels. A simple yet effective method has been designed to simulate low-quality patches with noises which conform to the distribution of real noises from human observers. To evaluate the effectiveness of our method, we have trained four state-of-the-art retinal vessel segmentation models using the purified dataset obtained from our method and compared their performance with those trained based on the original noisy datasets. Experimental results on two datasets DRIVE and CHASE\_DB1 demonstrate that obvious accuracy improvements can be achieved for all the four models when using the purified datasets from our method.

\end{abstract}
\section{Introduction}
Retinal vessel segmentation is a key step in screening, diagnosing and treatment of various ophthalmological diseases~\cite{fraz2012approach}. For instance, the tortuosity of segmented retinal vessels can characterize hypertensive retinopathy~\cite{miri2011retinal}. However, manually segmenting retinal vessels is tedious, time consuming and error-prone due to inadequate contrast between retinal vessels and unevenly distributed background, and a large variety in vessels’ shapes, sizes and intensities~\cite{fraz2012blood}.

A large number of methods~\cite{wu2018multiscale,yan2018joint} have been developed for automated retinal vessel segmentation. State-of-the-art methods mainly rely on supervised deep convolutional neural networks (CNNs) for their impressive results. To ensure a high segmentation accuracy, some methods focus on designing various task-specific constraints/priors which are then incorporated with CNNs for performance improvement. For instance, Fu~\emph{et al.}~\cite{fu2016deepvessel}  combined CNNs and conditional random field (CRF) to improve the spatial connectivity of vessel trees. Zhang~\emph{et al.}~\cite{zhang2018deep} utilized the edge-aware constraint and the deep supervision mechanism to improve the segmentation accuracy of tiny vessels and boundary regions. Other studies aimed at improving the quality of retinal images. For instance, Liskowski~\emph{et al.}~\cite{liskowski2016segmenting} investigated several pre-processing schemes, including global contrast normalization, zero-phase whitening, data augmentation via geometric transformations and gamma corrections. 

\begin{figure}[!t]
	\includegraphics[width=\textwidth]{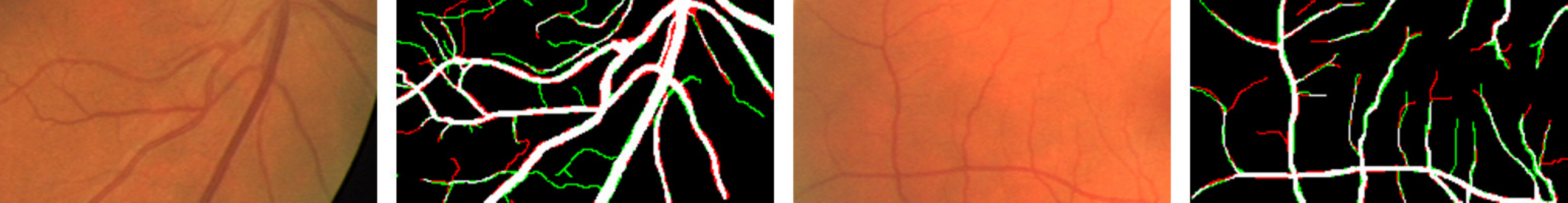}
	\caption{Illustration of inter-observer inconsistency among manual vessel labels. The white pixels denote common vessels annotated by the two observers, the red and green pixels denote vessels annotated only by the $1^{st}$ and $2^{nd}$ observer respectively.}
	\label{fig1}
\end{figure}
In addition to the capability of segmentation models and the quality of retinal images, the correctness of manual labels, which act as the ground truth to guide a model to learn vessel characteristics from raw image pixels, is also essential for the segmentation accuracy. However, several studies~\cite{yan2018skeletal,yan2018joint} have revealed that in most datasets there are nontrivial inter-observer inconsistency among manual labels. As shown in Fig.~\ref{fig1}, manually labeled vessels in different annotations don't completely match due to low contrast and limited resolution of fundus images: many thin vessels are missed in some annotations, and the boundary of thick vessels are incorrectly labeled. Noises arising from such inconsistent labels could severely mislead the training process~\cite{xiao2015learning} and in turn degrade the segmentation accuracy. On the other hand, collecting a large set of data with highly accurate annotations is infeasible due to the extremely high cost in the manual delineation of vessel trees by experienced experts and inevitable errors due to eye fatigue and insensitivity to visual characteristics of tiny retinal vessels. Automatically refining noisy manual labels is critical and effective for taking full advantage of existing datasets and improving the segmentation accuracy; however, few studies have been conducted along this direction so far.
  
\begin{figure}[!h]
	\includegraphics[width=\textwidth]{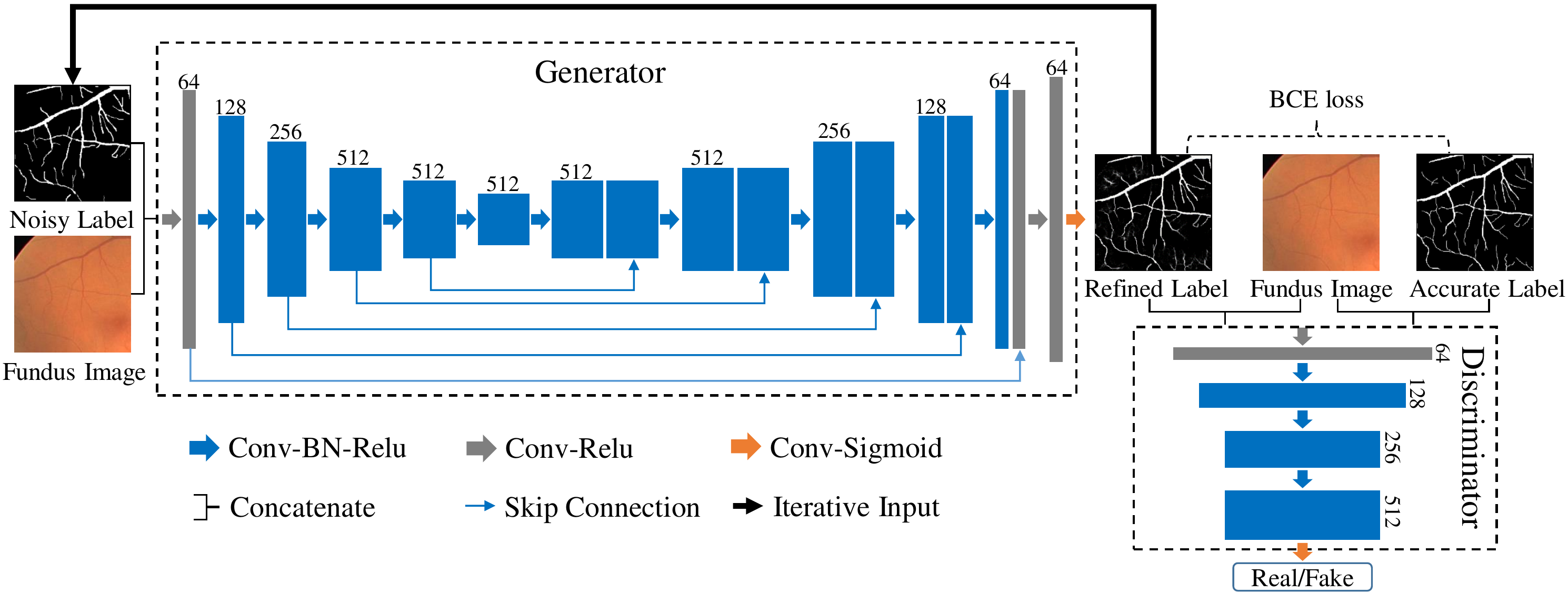}
	\caption{Framework of our iterative GAN for manual label refinement. } \label{fig2}
\end{figure}

To relieve the negative effect of the aforementioned annotation inconsistency or noise, we propose a novel label refinement method based on an iterative GAN, as shown in Fig.~\ref{fig2}. Our generator utilizes the U-Net as the backbone that takes a pair of retinal image and the corresponding noisy vessel labels as input and outputs a refined vessel label map with miss-labeled vessel pixels being added and noises being excluded. Such refinement process is repeated several times, in each iteration the refined vessel label map from the last round is used as the input of the generator. Our iterative generator is trained using both a pixel-wise binary cross-entropy (BCE) loss and an adversarial loss. The BCE loss ensures a high local similarity between a refined label map and the corresponding high-quality label map. The adversarial loss tries to guarantee the refined vessel label maps residing in the manifold of high-quality vessel label maps. 

Our training data is collected based on a set of high-quality retinal patches with consistent manual labels among different observers and low-quality retinal patches with noisy manual vessel labels. To obtain sufficient low-quality patches for training, we develop a method to simulate noisy vessel label maps. The simulated noises well conform to the distribution of real noises from human observers. We train four state-of-the-art retinal vessel segmentation models~\cite{ronneberger2015u,wu2018multiscale,zheng2015conditional} using the refined datasets and compared their performance with those trained based on the noisy datasets. Experimental results on two public datasets demonstrate that despite the high capability of existing segmentation models, improving the label purity of the training sets via our method can effectively yield obvious performance gain for vessel segmentation.

\section{Method}
This section details our label refinement method, which maps a noisy vessel label map to a more accurate label map based on an iterative GAN as shown in Fig.~\ref{fig2}. 
 
\subsection{Label Refinement with an Iterative GAN}
Given a retinal fundus image $x$ and the corresponding noisy manual vessel label map $z$, the label refinement model $G$ aims to generate a revised label map $G(x, z)$ which is closer to the high-quality label map $y$. To this end, we apply a conditional GAN framework~\cite{isola2017image}. For the generator G, we utilize the U-Net as the backbone for feature extraction and vessel label map decoding. The 3-channel retinal fundus image and the corresponding noisy vessel label map are directly concatenated to form a 4-channel input to the generator. 

To ensure the generated vessel maps reside in the manifold of the high-quality label maps and meanwhile share a high pixel-wise similarity, we train the generator using a combination of two losses, i.e. an adversarial loss from a discriminator and a pixel-wise binary cross-entropy (BCE) loss between the generated map and the high-quality map.

\textbf{Adversarial loss.} We utilize a PatchGAN discriminator to distinguish a pair of retinal image and the high-quality vessel label map $y$, i.e. $p_{real} = (x, y)$, from the a pair of retinal image and the generated vessel map, i.e. $p_{fake} = (x, G(x, z))$. Our discriminator consists of 5 convolutional layers with 4×4 kernels in each layer~\cite{isola2017image}. The adversarial loss is to maximize the Eq.~(\ref{eq2}):
\begin{equation}
	\mathcal{L}_{cGAN}(G,D) = \mathbb{E}_{x,y}[\log D(x,y)] + \mathbb{E}_{x,z}[\log(1 - D(x,G(x,z)))]
	\label{eq2}
\end{equation}

\textbf{Binary cross-entropy loss.} The BCE loss between a high-quality label map and the generated label map is calculated as
\begin{equation}
    \mathcal{L}_{BCE}(G) = \mathbb{E}_{x,y,z}[-\frac{1}{M}\sum_{j} (y_j\log(G(x,z)_j) + (1-y_j)\log(1-G(x,z)_j))]
\end{equation}
where $M$ is the number of pixels in a map and $j$ is the index of each pixel.
 
\begin{figure}[!t]
	\includegraphics[width=\textwidth]{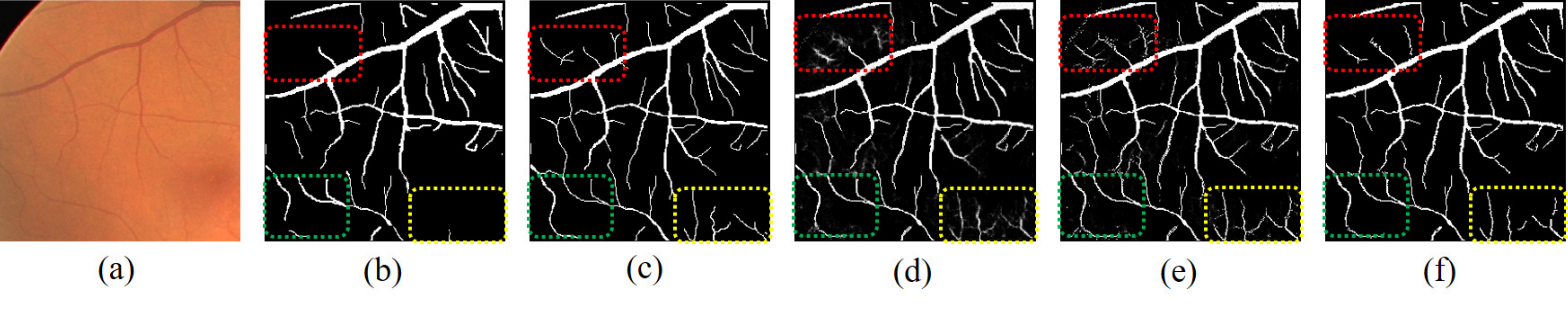}
	\caption{(a) A fundus image from DRIVE, (b) noisy vessel tree, (c) high-quality vessel tree, (d)-(e) refined vessel tree at the $1^{st}$ and $3^{rd}$ iteration respectively, (f) final refined vessel map after post-processing.}
	\label{fig3}
\end{figure}

Our refinement process is repeated several times, in each iteration the refined vessel label map from the last round is used as the input of the generator. The final objective is then formulated as
\begin{equation}
    G^* = arg\min_G\Big(\big(\max_D \sum_{i=1}^{N} w^{i}\mathcal{L}_{cGAN}^{i}(G, D)\big) + \lambda\sum_{i=1}^{N} w^{i}\mathcal{L}_{BCE}^{i}(G)\Big)
    \label{eq4}
\end{equation}
where $\lambda$ controls the importance of two losses,  $N$ is the number of iterations and $w$ denotes the weights of each iteration. Fig.~\ref{fig3}(b)-(e) illustrates the noisy label map, the high-quality label map and the refined noisy maps in the $1^{st}$ and $3^{rd}$ iterations. The  red and yellow boxes in Fig.~\ref{fig3} denote that miss-labeled vessels can be progressively recovered via our method and the green boxes indicate that the disconnected vessels can be re-connected and/or the boundary of thick vessels can be more accurately labeled via our method. We then apply post-processing based on connected component detection to exclude small spurious regions as noises, as shown in Fig.~\ref{fig3}(f).  

\subsection{Training Data Collection for the Label Refinement Model}
\textbf{High-quality vessel label map collection.} In this study, we denote high-quality labels as those consistent among different observers. For each vessel label map in the training set we randomly crop $N$ patches size of $256 \times 256$ from the original map ($565\times 584$), and choose those in which the ratio of vessel pixels is above 5\%. For each patch we assign a second observer to manually check the correctness of the annotated vessel pixels. If the intersection-over-union (IoU) between the original annotation and $2^{nd}$ annotation is greater than a certain threshold, we consider this patch as a high-quality label patch. 

\textbf{Noisy vessel label map simulation.} For each high-quality label patch, we simulate its noisy label patch for training. We observe that the main source of errors in manual annotations arises from miss-labeled thin vessels and incorrect indications of vessel boundaries. To simulate those noises, we apply the local erosion and/or dilation operations to the high-quality label patches to exclude some true thin vessels and/or change the thickness of vessels. In addition, we also apply the local open and/or close operations to the high-quality label maps to disconnect some vessels and/or merge some adjacent vessels, which usually occur in manual annotations due to the low image contrast. Given a high-quality label patch $v$, different types of noises could occur at different locations and we consider each type of noises appear locally in a small region. Therefore, we divide each patch $v$ regularly into $S$ grids $\{v_s | s\epsilon \{1, ..., S\}\}$ . We according to a certain probability choose one type of the four morphological operations for each grid $v_s$ and apply it to the grid to generate its simulated noisy version ${sim}(v_s)$,
\begin{small}
\begin{equation}
	     {sim}(v_s) = 
	     \begin{cases}
	      {Erode}(v_s)  &  \text{for } \quad 0  \leq p < 0.25 \\ 
	      {Dilate}(v_s)  &  \text{for } 0.25 \leq p < 0.50 \\
	      {Open}(v_s)  &  \text{for } 0.50 \leq p < 0.60 \\
	      {Close}(v_s)  &  \text{for } 0.60 \leq p < 0.70 \\
	      v_s  &  \text{otherwise.} \\
	    \end{cases}
\end{equation}
\end{small}
where $p$ is drawn from a uniform probability distribution \( \mathcal{U}(0, 1) \). \( {Erode} \) and \( {Dilate} \) are the erosion and dilation operators, specified by a square structuring element being set to 2. \( {Open} \) and \( {Close} \) are the opening and closing operators respectively, specified by a square structuring element being set to 3. Once $v_s$ has been artificially degraded, it is stored back at its original location in $v$. Examples of simulated maps are provided in supplementary material.

\section{Experiments and Results}
\subsection{Dataset}
We evaluate our methods using two datasets DRIVE~\cite{staal2004ridge} and CHASE\_DB1~\cite{owen2009measuring}. DRIVE consists of 40 fundus images size of $565 \times 584$. Each training image has one manual annotation from a single observer. The dataset is equally divided into training and test set. CHASE\_DB1 contains 28 fundus images size of  $999 \times 960$. Each image has two manual annotations from two observers. The dataset is split into 20 images for training and 8 images for testing as consistent with previous literatures~\cite{yan2018joint}. 

To train the label refinement network, we collected the high-quality patches and simulated noisy patches from the 20 training images of DRIVE for its high quality in annotations. We assigned a $2^{nd}$ observer to manually check the correctness of every annotated pixels in the $1^{st}$ annotatons. For each training image, we chose 300 high-quality patches and thus yield in total 6000 original high-quality vessel label patches. For each high-quality patch, we simulate a noisy version and finally yield a total of 6000 training pairs. 

\subsection{Label Refinement Model Training Details}
To train our label refinement model, in Eq.~(\ref{eq4}), we set the weight $\lambda$ to 50, the number of iterations $N$ to 3 and the weights of each iteration $w_i$ to 1.0, 1.6, 2.2 respectively. To optimize our networks, we follow the standard approach from~\cite{isola2017image}. Specifically, we alternate between one gradient descent step in $D$, then one step in $G$. We use mini-batch SGD and apply the Adam solver, with a learning rate of 0.0002, and momentum parameters $\beta_1= 0.5$, $\beta_2= 0.999$.

\subsection{Segmentation Models}
We compared the segmentation model trained based on the original manual labels vs. the model trained based on the refined manual labels via our method. We choose four state-of-the-art segmentation methods, i.e. U-Net~\cite{ronneberger2015u}, U-Net+CRF~\cite{zheng2015conditional}, MS and MS-NFN~\cite{wu2018multiscale} due to their high capability in vessel segmentation.

For each segmentation model, we used the training images (with original labels or refined labels via our method) in the dataset for training. We extracted 190K patches size of $48 \times 48$ by randomly cropping 9500 overlapping patches in each of the 20 DRIVE training images, and 90K patches of size $64 \times 64$ from the CHASE\_DB1. We evaluate the segmentation accuracy using the test images with the original labels for a fair comparison with the state-of-the-art methods. We followed the same evaluate metrics, i.e. accuracy (Acc), sensitivity (Se), specificity (Sp) and area under curve (AUC), which are widely used for retinal vessel segmentation~\cite{wu2018multiscale,zhang2018deep}.

\subsection{Results}
We follow the practice of using the $1^{st}$ observer's annotation as ground truth for evaluation and apply the Otsu~\cite{otsu1979threshold} algorithm to automatically convert a probability map to a binary segmentation map. Results in Tables~\ref{tab1} and~\ref{tab2} show that obvious improvements can be obtained when using the refinement labels compared with the original manual labels for all methods. These results demonstrate that despite high capability of existing segmentation models, refining manual labels for training can yield nontrivial gains.  

Moreover, significant improvements of performance are observed on CHASE\_DB1 though our label refinement network is trained using high-quality patches from DRIVE only, which demonstrates the vitality of our proposed strategy of high-quilty patch selection and noisy label simulation, and the promising generalization of the label refinement network.
\begin{table}[!t]
	\caption{Performance comparison of different labels on CHASE\_DB1.}\label{tab1}
	\begin{center}
		\begin{tabular}{c|c c c c|c c c c}
			\hline
			\multirow{2}{*}{Ground Truth} & \multicolumn{4}{c}{U-Net} & \multicolumn{4}{|c}{U-Net-CRF}  \\
			\cline{2-9}
			& Acc & Sp & Se & Auc & Acc & Sp & Se & Auc \\
			\hline
			$1^{st}$ original & 0.9627 & 0.9788 &  0.8014 & 0.9798 & 0.9629 &  0.9806 & 0.7864 &  0.9814 \\
			\hline
			$1^{st}$ refined & 0.9622 & \textbf{0.9821} &  0.7625 & \textbf{0.9806} & \textbf{0.9633} &  0.9785 & \textbf{0.8101} &  \textbf{0.9821} \\
			\hline
			\hline
			$2^{nd}$ original & 0.9606 & 0.9765 &  0.8023 & 0.9794 & 0.9620 &  0.9796 & 0.7859 &  0.9806 \\
			\hline
			$2^{nd}$ refined & \textbf{0.9614} & \textbf{0.9766} &  \textbf{0.8079} & \textbf{0.9801} & \textbf{0.9630} &  \textbf{0.9809} & 0.7842 &  \textbf{0.9815} \\
			\hline
			\hline
			\multirow{2}{*}{Ground Truth} & \multicolumn{4}{c}{MS} & \multicolumn{4}{|c}{MS-NFN}  \\
			\cline{2-9}
			& Acc & Sp & Se & Auc & Acc & Sp & Se & Auc \\
			\hline
			$1^{st}$ original & 0.9608 & 0.9788 &  0.7813 & 0.9788 & 0.9631 &  0.9805 & 0.7880 &  0.9817 \\
			\hline
			$1^{st}$ refined & \textbf{0.9612} & 0.9774 &  \textbf{0.7979} & \textbf{0.9798} & \textbf{0.9634} &  0.9805 & \textbf{0.7920} &  \textbf{0.9823} \\
			\hline
			\hline
			$2^{nd}$ original & 0.9589 & 0.9750 &  0.7978 & 0.9772 & 0.9607 &  0.9736 & 0.8315 &  0.9812 \\
			\hline
			$2^{nd}$ refined & \textbf{0.9599} & \textbf{0.9755} &  \textbf{0.8036} & \textbf{0.9783} & \textbf{0.9609} &  0.9724 & \textbf{0.8458} &  \textbf{0.9819} \\
			\hline
		\end{tabular}
	\end{center}
\end{table}

\begin{table}[!t]
	\caption{Performance comparison of different labels on DRIVE.}\label{tab2}
	\begin{center}
		\begin{tabular}{c|c c c c|c c c c}
			\hline
			\multirow{2}{*}{Ground Truth} & \multicolumn{4}{c}{U-Net} & \multicolumn{4}{|c}{U-Net-CRF} \\
			\cline{2-9}
			& Acc & Sp & Se & Auc & Acc & Sp & Se & Auc\\
			\hline
			$1^{st}$ original & 0.9547 & 0.9856 & 0.7435 & 0.9784 & 0.9563 &  0.9824 & 0.7771 &  0.9794 \\
			\hline
			$1^{st}$ refined & \textbf{0.9551} & 0.9854 &  \textbf{0.7469} & \textbf{0.9786} & 0.9562 &  \textbf{0.9831} & 0.7713 & \textbf{0.9796} \\
			\hline
			\hline
			\multirow{2}{*}{Ground Truth} & \multicolumn{4}{c}{MS} & \multicolumn{4}{|c}{MS-NFN} \\
			\cline{2-9}
			& Acc & Sp & Se & Auc & Acc & Sp & Se & Auc \\
			\hline
			$1^{st}$ original & 0.9562  & 0.9799 &  0.7934 & 0.9798 & 0.9566  & 0.9780 &  0.8100 & 0.9806 \\
			\hline
			$1^{st}$ refined & \textbf{0.9565} & \textbf{0.9803} &  0.7930 & \textbf{0.9801} & \textbf{0.9567}  & \textbf{0.9802} &  0.7952 & \textbf{0.9807}\\
			\hline
		\end{tabular}
	\end{center}
\end{table}
\section{Conclusions}
In this paper, we bring up the inconsistency problem in manual annotations of medical image and introduce a solution to refine the labels for accuracy improvement. Our label refinement can be used for helping any off-the-shelf segmentation method and bring nontrivial gains in the segmentation accuracy.

%
%
%
%
%
\bibliographystyle{splncs04}
\bibliography{mybibliography}

\begin{thebibliography}{10}
\providecommand{\url}[1]{\texttt{#1}}
\providecommand{\urlprefix}{URL }
\providecommand{\doi}[1]{https://doi.org/#1}

\bibitem{fraz2012approach}
Fraz, M.M., et~al.: An approach to localize the retinal blood vessels using bit
  planes and centerline detection. Computer methods and programs in biomedicine
   \textbf{108}(2),  600--616 (2012)

\bibitem{fraz2012blood}
Fraz, M.M., et~al.: Blood vessel segmentation methodologies in retinal
  images--a survey. Computer methods and programs in biomedicine
  \textbf{108}(1),  407--433 (2012)

\bibitem{fu2016deepvessel}
Fu, H., Xu, Y., Lin, S., Wong, D.W.K., Liu, J.: Deepvessel: Retinal vessel
  segmentation via deep learning and conditional random field. In:
  International Conference on Medical Image Computing and Computer-Assisted
  Intervention. pp. 132--139. Springer (2016)

\bibitem{NIPS2014_5423}
Goodfellow, I., et~al.: Generative adversarial nets. Advances in Neural
  Information Processing Systems 27  (2014)

\bibitem{isola2017image}
Isola, P., Zhu, J.Y., Zhou, T., Efros, A.A.: Image-to-image translation with
  conditional adversarial networks. In: Proceedings of the IEEE conference on
  computer vision and pattern recognition. pp. 1125--1134 (2017)

\bibitem{liskowski2016segmenting}
Liskowski, P., Krawiec, K.: Segmenting retinal blood vessels with deep neural
  networks. IEEE transactions on medical imaging  \textbf{35}(11),  2369--2380
  (2016)

\bibitem{miri2011retinal}
Miri, M.S., Mahloojifar, A.: Retinal image analysis using curvelet transform
  and multistructure elements morphology by reconstruction. IEEE Transactions
  on Biomedical Engineering  \textbf{58}(5),  1183--1192 (2011)

\bibitem{otsu1979threshold}
Otsu, N.: A threshold selection method from gray-level histograms. IEEE
  transactions on systems, man, and cybernetics  \textbf{9}(1),  62--66 (1979)

\bibitem{owen2009measuring}
Owen, C.G., Rudnicka, A.R., Mullen, R., Barman, S.A., Monekosso, D., Whincup,
  P.H., Ng, J., Paterson, C.: Measuring retinal vessel tortuosity in
  10-year-old children: validation of the computer-assisted image analysis of
  the retina (caiar) program. Investigative ophthalmology \& visual science
  \textbf{50}(5),  2004--2010 (2009)

\bibitem{ronneberger2015u}
Ronneberger, O., Fischer, P., Brox, T.: U-net: Convolutional networks for
  biomedical image segmentation. In: International Conference on Medical image
  computing and computer-assisted intervention. pp. 234--241. Springer (2015)

\bibitem{staal2004ridge}
Staal, J., Abr{\`a}moff, M.D., Niemeijer, M., Viergever, M.A., Van~Ginneken,
  B.: Ridge-based vessel segmentation in color images of the retina. IEEE
  transactions on medical imaging  \textbf{23}(4),  501--509 (2004)

\bibitem{wu2018multiscale}
Wu, Y., Xia, Y., Song, Y., Zhang, Y., Cai, W.: Multiscale network followed
  network model for retinal vessel segmentation. In: International Conference
  on Medical Image Computing and Computer-Assisted Intervention. pp. 119--126.
  Springer (2018)

\bibitem{xiao2015learning}
Xiao, T., Xia, T., Yang, Y., Huang, C., Wang, X.: Learning from massive noisy
  labeled data for image classification. In: Proceedings of the IEEE Conference
  on Computer Vision and Pattern Recognition. pp. 2691--2699 (2015)

\bibitem{yan2018skeletal}
Yan, Z., Yang, X., Cheng, K.T.: A skeletal similarity metric for quality
  evaluation of retinal vessel segmentation. IEEE transactions on medical
  imaging  \textbf{37}(4),  1045--1057 (2018)

\bibitem{yan2018joint}
Yan, Z., Yang, X., Cheng, K.T.T.: Joint segment-level and pixel-wise losses for
  deep learning based retinal vessel segmentation. IEEE Transactions on
  Biomedical Engineering  (2018)

\bibitem{zhang2018deep}
Zhang, Y., Chung, A.C.: Deep supervision with additional labels for retinal
  vessel segmentation task. In: International Conference on Medical Image
  Computing and Computer-Assisted Intervention. pp. 83--91. Springer (2018)

\bibitem{zheng2015conditional}
Zheng, S., Jayasumana, S., Romera-Paredes, B., Vineet, V., Su, Z., Du, D.,
  Huang, C., Torr, P.H.: Conditional random fields as recurrent neural
  networks. In: Proceedings of the IEEE international conference on computer
  vision. pp. 1529--1537 (2015)

\end{thebibliography}

\end{document}